\documentclass[12]{iopart}
\usepackage{iopams}
\usepackage{amssymb}
\usepackage{amsfonts}
\usepackage{epsfig}

\begin{document}
\title{Berry phase and entanglement of 3 qubits in a new Yang-Baxter system  }

\author{Taotao Hu}
\address{Department of Physics,
Northeast Normal University, Changchun, Jilin 130024, People's
Republic of China}
\author{Kang Xue}
\ead{Xuekang@nenu.edu.cn}\address{Department of Physics, Northeast Normal University,
Changchun, Jilin 130024, People's Republic of China}
\author{Chunfeng Wu}
\address{Department of Physics, National University of
singapore, 2 Science Drive 3, singapore 117542} \address{Centre for
Quantum Technologies, National University of singapore, 3 Science
Drive 2, singapore 117543}

\begin{abstract}
In this paper we construct a new $8\times8$ $\mathbb{M}$ matrix from
the $4\times4$ $M$ matrix, where $\mathbb{M}$/$M$ is the image of the braid group
representation. The $ 8\times8 $ $\mathbb{M}$ matrix and the
$4\times4$ $M$ matrix both satisfy extraspecial 2-groups algebra
relations. By Yang-Baxteration approach, we derive a unitary $
\breve{R}(\theta , \varphi)$ matrix from the $\mathbb{M}$ matrix with
parameters $\varphi$ and $\theta$. Three-qubit entangled
states can be generated by using the $\breve{R}(\theta,\varphi)$
matrix. A Hamiltonian for 3 qubits is constructed from the unitary
$\breve{R}(\theta,\varphi)$ matrix. We then study the entanglement
and Berry phase of the Yang-Baxter system.
\end{abstract}


\maketitle

\section{Introduction}

Recently, it has been revealed that there are natural and profound
connections between quantum information theory and braid group
theory \cite{H.A,L.H,Y.Z,Y.Z.L,J.F,M.N,C.N,R.J}. In a recent paper
by Franko, Rowell and Wang \cite{J.F}, the images of the unitary
braid group representations generated by the original $4\times4$
Bell matrix have been identified as extensions of extraspecial
2-groups \cite{Y.Z.E,Yging}. Extraspecial 2-groups are now known to
play an interesting role in theory of quantum information,
particularly in the theory of quantum error correction. They provide
a bridge between quantum error correcting codes and binary
orthogonal geometry \cite{A.C}. Moreover, they form a subgroup of
the pauli group, which is of importance in the theory of stablized
code \cite{A.C.E,D.G}. In Ref. \cite{J.F}, the authors utilized
extraspecial 2-groups to study the images of braid group $ B_{n}$
under the representation associated with the $ 4\times4$ Bell
matrix. It is found that the extraspecial 2-groups  are the central
link between almost-complex structures and unitary braid
representations. New higher dimensional unitary braid group
representations can be constructed by considering an extension of
the extraspecial 2-groups \cite{Y.Z.E}.

Many efforts have been devoted to investigate the braiding operators
and Yang-Baxter eqauation (YBE)
to the field of quantum information and quantum computation. This
provides a novel way to study the quantum entanglement
\cite{J.L.K,cxg1,cxg,ckg2}. It is shown that braiding transformation
is a natural approach describing quantum entanglement by applying
the unitary braiding operators to realize entanglement swapping and
generate the GHZ states as well as the linear cluster states.
Unitary solutions of braid group relation \cite{H.A,L.H} and the YBE
\cite{Y.Z,Y.Z.L} can be identified with universal quantum gates
\cite{J.L}. The investigation of many types of Yang-Baxter systems
have attracted researchers' attention. Usually, a Hamiltonian can be
constructed from the unitary $\breve{R}(\theta,\varphi)$ matrix by
Yang-Baxterization approach. Yang-Baxterization \cite{V.F.R,M.L} is
exploited \cite{Y.Z,Y.Z.L} to derive the Hamiltonian for the unitary
evolution of entangled states.

In this paper, we present a new $8\times8$ $\mathbb{M}$ matrix from
the $4\times4$ $M$ matrix, where $\mathbb{M}$/$M$ is the image of braid group
representation. In Sec \ref{sec2}, we briefly review the Yang-Baxter
systems of $4\times4$ $M$ matrix. Then we construct a new $8\times8$
$\mathbb{M}$ matrix, which satisfies the extraspecial 2-groups
algebra relation, from the $4\times4$ $M$ matrix. By using
Yang-Baxterization approach, we derive a unitary
$\breve{R}(\theta,\varphi)$ matrix based on the $8\times8$
$\mathbb{M}$-matrix. We analyze the effect of
$\breve{R}(\theta,\varphi)$ matrix on the entanglement of three
qubits. In Sec \ref{sec3}, we construct a Hamiltonian of 3 qubits
from the unitary $\breve{R}(\theta,\varphi)$ matrix and investigate
the entanglement and Berry phase of this system. The results are
summarized in the last section.

\section{a new $ 8\times8 $ $\mathbb{M}$ matrix and its Yang-Baxterization in 3-qubits system}\label{sec2}
For $ 4\times4$ Yang-Baxter systems of 2 qubits, we know that the
rational solution of the YBE can be expressed as
$\breve{R}_{i,i+1}(\theta,\varphi) =\sin \theta I_{i}\otimes I_{i+1}
+  \cos \theta M_{i,i+1}$\cite{J.L.K}, where $\theta$ is the
spectral parameter and $M_{i,i+1}$ is an operator of braid group
which can be expressed in the form of spin operators
$S^{+}_{i}=S^{1}_{i}+i S^{2}_{i}$ and $S^{-}_{i}=S^{1}_{i}-i
S^{2}_{i}$ as follows
\begin{eqnarray}\label{M}
M_{i,i+1}=
e^{-i\varphi}S_{i}^{+}S_{i+1}^{+}-e^{i\varphi}S_{i}^{-}S_{i+1}^{-}+S_{i}^{+}S_{i+1}^{-}-S_{i}^{-}S_{i+1}^{+}.
\end{eqnarray}
The operator $M_{i,i+1}$ satisfies
the extraspecial 2-groups relations
\cite{Yging}:
\begin{eqnarray}
M_{i,i+1}^{2}&=&\alpha I \nonumber\\
M_{i,i+1}M_{i+1,i+2}M_{i,i+1}&=&M_{i+1,i+2} \nonumber\\
M_{i+1,i+2}M_{i,i+1}M_{i+1,i+2}&=&M_{i,i+1}.
\end{eqnarray}
In terms of the standard basis $\{|00\rangle,|01
\rangle,|10\rangle, |11\rangle\}$ of 2 qubits, the $M_{i,i+1}$ matrix is
of the following form:
\begin{eqnarray}
 M_{i,i+1}=\left(
  \begin{array}{cccc}
    0 & 0  & 0 & e^{-i\varphi} \\
    0 & 0 & 1 & 0 \\
   0  & 1 & 0 & 0 \\
    -e^{i \varphi } & 0 & 0 & 1 \\
  \end{array}
\right).
\end{eqnarray}
When the unitary matrix $\breve{R}_{i,i+1}(\theta,
\varphi)$=$\sin\theta I_{i}\otimes I_{i+1} + i \cos \theta
M_{i,i+1}$ acts on the standard basis of 2 qubits, one gets
four entangled states which possess the same entanglement degree of
$|\sin2\theta|$ \cite{J.L.K}.

In the following, we construct a $ 8\times8 $ matrix $\mathcal {M}$ from
this $ 4\times4 $ matrix M. In order to connect three qubits, we write $\mathcal {M}$
by using $M_{j,j+1}\;(j=i,i+1)$
\begin{eqnarray}\label{m2}
\mathcal {M}=\frac{1}{\sqrt{3}}(M_{i,i+1}\otimes
I_{i+2}+I_{i}\otimes M_{i+1,i+2}+M_{i+1,i+2}M_{i,i+1}).
\end{eqnarray}
$\mathcal {M}$ can be expressed by using spin operators:
\begin{eqnarray}\label{mt}
 \mathcal {M} =\frac{1}{\sqrt{3}}
[e^{-i\varphi}(S_{i}^{+}S_{i+1}^{+}+S_{i+1}^{+}S_{i+2}^{+})-
e^{i\varphi}(S_{i}^{-}S_{i+2}^{-}+S_{i+1}^{-}S_{i+2}^{-})\nonumber~~~\\
+(S_{i}^{+}S_{i+1}^{-}+S_{i+1}^{+}S_{i+2}^{-}-S_{i}^{-}S_{i+1}^{+}-S_{i+1}^{-}S_{i+2}^{+})\nonumber~~~\\
2S_{i+1}^{3}(e^{-i\varphi}S_{i}^{+}S_{i+2}^{+}-e^{i\varphi}S_{i}^{-}S_{i+2}^{-}+S_{i}^{+}S_{i+2}^{-}-S_{i}^{-}S_{i+2}^{+})]
\end{eqnarray}
Write in terms of the basis of three qubits $\{|000\rangle$, $|001\rangle$,
$|010\rangle$, $|011\rangle$, $|100\rangle$, $|101\rangle$, $|110\rangle$,
$|111\rangle\}$, the $ 8\times8 $ matrix $\mathcal{M}$ is
\begin{eqnarray}
\mathcal {M}=\frac{1}{\sqrt{3}}\left(
\begin{array}{cccccccc}
0 & 0 & 0 & e^{-i\varphi}& 0&e^{-i\varphi}&e^{-i\varphi}& 0\\
0 & 0 & 1 & 0& 1& 0& 0&e^{-i\varphi}\\
0 & -1 & 0 & 0& 1& 0& 0& -e^{-i\varphi} \\
-e^{i\varphi} & 0 & 0 & 0& 0& 1& -1& 0\\
0 & -1 & -1 & 0& 0& 0& 0& e^{-i\varphi} \\
-e^{i\varphi}& 0 & 0 & -1& 0& 0& 1& 0\\
-e^{i\varphi} & 0 & 0 & 1& 0& -1& 0& 0 \\
0 & -e^{i\varphi} & e^{i\varphi} & 0& -e^{i\varphi}& 0& 0& 0\\
\end{array}
\right).
\end{eqnarray}
Introduce $\mathbb{M}=-i\mathcal {M}$, it is not difficult to find that
$\mathbb{M}$ satisfies the following extraspecial 2-groups relations
\cite{Yging}:
\begin{eqnarray}
&&\mathbb{M}^{2} = \alpha I \nonumber\\
&&\mathbb{M}_{12}^{\frac{3}{2}\frac{1}{2}}\mathbb{M}_{23}^{\frac{1}{2}\frac{3}{2}}\mathbb{M}
_{12}^{\frac{3}{2}\frac{1}{2}}=\mathbb{M}_{12}^{\frac{3}{2}\frac{1}{2}}\nonumber\\
&&\mathbb{M}_{23}^{\frac{1}{2}\frac{3}{2}}\mathbb{M}_{12}^{\frac{3}{2}\frac{1}{2}}\mathbb{M}_{23}^{\frac{1}{2}\frac{3}{2}}=\mathbb{M}_{23}^{\frac{1}{2}\frac{3}{2}}
\end{eqnarray}

It is known that a unitary solution of the YBE can be found via Yang-Baxterization on
the solution of the extraspecial 2-groups relations.
The Yang-Baxterization of the extraspecial 2-group operator
$\mathbb{M}$ is \cite{J.L.K}:
\begin{equation}\label{m}
\breve{R}(x)=\rho(x) (\mathcal {I} + G(x) \mathbb{M}).
\end{equation}
where $\rho(x)$ and G(x) are some functions of $x$ to be determined,
$\mathcal {I}=I_{i}\otimes I_{i+1}\otimes I_{i+2}$ is identity matrix.
One can choose appropriate $\rho(x)$ and $G(x)$ so that
$\breve{R}(x)$ is unitary. The unitary $\breve{R}$-matrix satisfies the YBE which is of the form,
\begin{equation}\label{ybe}
\breve{R}_{12}(x)\breve{R}_{23}(xy)\breve{R}_{12}(y)=\breve{R}_{23}(y)\breve{R}_{12}(xy)\breve{R}_{23}(x),
\end{equation}
where multiplicative parameters $x$ and $y$ are known as the spectral
parameters. In order to make $\breve{R}(x)$ a unitary matrix,
$\breve{R}^{\dag}(x)$ should be equal to the inverse $\breve{R}^{-1}(x)$.
In this way, we obtain that $\breve{R}(x)
=\frac{x+x^{-1}}{2}(\mathcal {I} +
\frac{x-x^{-1}}{x+x^{-1}}\mathbb{M})$. By introducing a
new variable parameter $\theta$ as $\frac{x-x^{-1}}{2} =i\cos
\theta $ and $ \frac{x+x^{-1}}{2}=\sin \theta $, the matrix
$\breve{R}(x)$ can be rewritten as $\breve{R}(\theta , \varphi) $=$ \sin
\theta \mathcal {I} + i \cos \theta \mathbb{M}$ = $\sin\theta \mathcal
{I} + \cos\theta \mathcal {M}$. In the following, we express $\breve{R}(\theta , \varphi)$
in the form of spin operators
\begin{eqnarray}
\breve{R}(\theta , \varphi)=\sin\theta I_{i}\otimes I_{i+1}\otimes
I_{i+2}+\frac{1}{\sqrt{3}}\cos\theta(e^{-i\varphi}(S_{i}^{+}S_{i+1}^{+}+S_{i+1}^{+}S_{i+2}^{+})\nonumber~~~\\
-e^{i\varphi}(S_{i}^{-}S_{i+1}^{-}+S_{i+1}^{-}S_{i+2}^{-})
+(S_{i}^{+}S_{i+1}^{-}+S_{i+1}^{+}S_{i+2}^{-}-S_{i}^{-}S_{i+1}^{+}-S_{i+1}^{-}S_{i+2}^{+})
\nonumber~~~\\
+2S_{i+1}^{3}(e^{-i\varphi}S_{i}^{+}S_{i+2}^{+}-e^{i\varphi}S_{i}^{-}S_{i+2}^{-}+S_{i}^{+}S_{i+2}^{-}-
S_{i}^{-}S_{i+2}^{+})).
\end{eqnarray}
 When the unitary matrix $\breve{R}(\theta , \varphi)$ acts on the
direct product states
$|klm\rangle\equiv|k\rangle_i\otimes|l\rangle_{i+1}\otimes
|m\rangle_{i+2}$, the $\breve{R}(\theta ,\varphi)$ matrix transfers
product states to entangled states
\begin{eqnarray}\label{tai}
|000\rangle\rightarrow\sin \theta |000\rangle-\frac{1}{\sqrt{3}} \cos \theta e^{i \varphi}|011\rangle
-\frac{1}{\sqrt{3}} \cos\theta e^{i \varphi}|101\rangle-\frac{1}{\sqrt{3}}\cos\theta e^{i \varphi}|110\rangle, \nonumber \\
|001\rangle\rightarrow\sin \theta |001\rangle-\frac{1}{\sqrt{3}} \cos \theta|010\rangle
-\frac{1}{\sqrt{3}} \cos\theta|100\rangle-\frac{1}{\sqrt{3}}\cos\theta e^{i \varphi}|111\rangle, \nonumber \\
|010\rangle\rightarrow\sin \theta |010\rangle+\frac{1}{\sqrt{3}} \cos \theta|001\rangle
-\frac{1}{\sqrt{3}} \cos\theta|100\rangle+\frac{1}{\sqrt{3}}\cos\theta e^{i \varphi}|111\rangle, \nonumber \\
|011\rangle\rightarrow\sin \theta |011\rangle+\frac{1}{\sqrt{3}} \cos \theta e^{-i \varphi}|000\rangle
-\frac{1}{\sqrt{3}} \cos\theta|101\rangle+\frac{1}{\sqrt{3}}\cos\theta|110\rangle, \nonumber \\
|100\rangle\rightarrow\sin \theta |100\rangle+\frac{1}{\sqrt{3}} \cos \theta |001\rangle
+\frac{1}{\sqrt{3}} \cos\theta|010\rangle-\frac{1}{\sqrt{3}}\cos\theta e^{i \varphi}|111\rangle, \nonumber \\
|101\rangle\rightarrow\sin \theta |101\rangle+\frac{1}{\sqrt{3}} \cos \theta e^{-i \varphi}|000\rangle
+\frac{1}{\sqrt{3}} \cos\theta |011\rangle-\frac{1}{\sqrt{3}}\cos\theta|110\rangle, \nonumber \\
|110\rangle\rightarrow\sin \theta |110\rangle+\frac{1}{\sqrt{3}} \cos \theta e^{-i \varphi}|000\rangle
-\frac{1}{\sqrt{3}} \cos\theta|011\rangle+\frac{1}{\sqrt{3}}\cos\theta |101\rangle, \nonumber \\
|111\rangle\rightarrow\sin \theta |111\rangle+\frac{1}{\sqrt{3}} \cos \theta e^{-i \varphi}|001\rangle
-\frac{1}{\sqrt{3}} \cos\theta e^{-i \varphi}|010\rangle+\frac{1}{\sqrt{3}}\cos\theta e^{-i \varphi}|100\rangle, \nonumber \\
\end{eqnarray}
For convenience we label the $i$-th, $(i+1)$-th, and $(i+2)$-th qubits by A, B, and C respectively.
Tripartite entangled states can be measured by three-tangle (or residual tangle) $\tau_{ABC}$
proposed by Coffman, Kudu, and Wootters \cite{C.K.W}. $\tau_{ABC}$ is expressed by using the composing
coefficients $a_{ijk}$ corresponding to the basis state $|ijk\rangle$,
\begin{eqnarray}\label{t}
\tau &=& 4|d_{1}-2d_{2}+4d_{3}| \nonumber\\
d_{1}&=& a_{000}^{2}a_{111}^{2}+a_{001}^{2}a_{110}^{2}+a_{010}^{2}a_{101}^{2}+a_{100}^{2}a_{011}^{2} \nonumber\\
d_{2}&=& a_{000}a_{111}a_{011}a_{100}+a_{000}a_{111}a_{101}a_{010}+a_{000}a_{111}a_{110}a_{001}\nonumber\\
&+&a_{011}a_{100}a_{101}a_{010}+a_{011}a_{100}a_{110}a_{001}+a_{101}a_{010}a_{110}a_{001}\nonumber\\
d_{2}&=& a_{000}a_{110}a_{101}a_{011}+a_{111}a_{001}a_{010}a_{100}
\end{eqnarray}
The above eight 3-qubit entangled
states are found to have the same entanglement degree as
\begin{eqnarray}\label{t1}
\tau_{ABC}= \frac{16\sqrt{3}|\sin\theta \cos^{3}\theta|}{9}.
 \end{eqnarray}

The concurrunce \cite{W.K.W} measuring bipartite entangled states is defined as
\begin{eqnarray}\label{con}
C(\rho)= {\rm
max}\{0,\lambda_{1}-\lambda_{2}-\lambda_{3}-\lambda_{4}\}
 \end{eqnarray}
where  $\{\lambda_{i}\}$ are the eigenvalues of the matrix
$\rho_{12}(\sigma_{y}^{1}\otimes
\sigma_{y}^{2})\rho^{\ast}_{12}(\sigma_{y}^{1}\otimes
\sigma_{y}^{2})$, here $\rho_{12}$ is the density matrix of the pair
and it can be either pure or mixed  with $\rho^{\ast}_{12}$ denoting
the complex conjugate of $\rho$ and $\sigma_{y}^{1/2}$ are the Pauli
matrices for atoms 1 and 2. This quantity attains its maximum value
of 1 for maximally entangled states and vanishes for separable
states.

For the 2-qubit subsystem of three qubits, we need consider the
reduced density. i.e. $\rho_{AB}=Tr_{c}\rho$, here $\rho$ is the
density of 3 qubits. It is pure and has the form
$\rho=|\psi\rangle\langle\psi|$. Though calculating the reduced
density of the eight 3 qubits states we get in  Eq (\ref{tai}), it
is not difficult to obtain the corresponding concurrences as follows
\begin{eqnarray}\label{c2}
C_{AB}=C_{BC}=C_{AC} = |\frac{1}{\sqrt{3}}|\sin
2\theta|-\frac{2}{3}\cos^{2}\theta|.
\end{eqnarray}
It is worth noting that the eight 3 qubits state we get in  Eq
(\ref{tai}) all have the same pairwise concurrence as  Eq
(\ref{c2}). On the other hand, the entanglement between qubit A and
the pair BC can be calculated \cite{C.K.W} as
\begin{eqnarray}\label{c3}
C_{A(BC)}^{2}=C_{BC}^{2}+C_{AC}^{2}+\tau_{ABC}.
 \end{eqnarray}
Thus we have
\begin{eqnarray}\label{c4}
 C_{A(BC)}^{2} =\frac{8}{9}\cos^{2}\theta(1+ 2\sin^{2}\theta).
 \end{eqnarray}
One can easily verfiy that $C_{B(AC)}^{2}= C_{C(AB)}^{2}=
 C_{A(BC)}^{2}$. So it is worth noting that by this $8\times8$ $\breve{R}(\theta , \varphi)$ acting on the product states
of 3 qubits, one can also get eight tripartite entangled states with
the same degree of entanglement. While for the $4\times4$
Yang-Baxter system, they get the same entanglement degree of 2
qubits $C_{i,i+1}=|\sin2\theta|$. From Eqs (\ref{t1}), (\ref{c2})
and (\ref{c4}), one can see that when $\theta=\frac{\pi}{6}$
$C_{AB}=C_{BC}=C_{AC} =0$ and $C_{B(AC)}^{2}=
C_{C(AB)}^{2}=C_{A(BC)}^{2}$= $\tau_{ABC}$=1. This case is just the
GHZ state \cite{C.K.W}. According to Eq. (\ref{t1}), when
 $\cos\theta=0$, the three-tangle $\tau_{ABC}$ is
equal to 0 and $C_{AB}=C_{BC}=C_{AC}=0$. This tells us that the
states in Eq. (\ref{tai}) are separable in this case. When
$\sin\theta$=0, $C_{A(BC)}^{2}=C_{BC}^{2}+C_{AC}^{2}$ while
$\tau_{ABC}$=0. This corresponds to the W state \cite{C.K.W}. So one
can see the  $8\times8$ $\breve{R}(\theta , \varphi)$ generates
arbitrary tripartite entangled states  which can be achieved
depending on the parameters $\theta$.

\section{Hamiltonian, Entanglement and Berry phase}\label{sec3}
Generally, multi-spin interaction Hamiltonians can be constructed based on the YBE. As $\breve{R}$ is unitary,
it can define the evolution of a state $|\Psi(0)\rangle$
\begin{eqnarray}
|\Psi(t)\rangle=\breve{R}_i(t)|\Psi(0)\rangle,
\end{eqnarray}
here $\breve{R}_i(t)$ is time-dependent, which can be realized by specifying corresponding
time-dependent parameter of $\breve{R}_i$.
By taking partial derivative of the state $|\Psi(t)\rangle$ with respect to time $t$, we have an equation
\begin{eqnarray}
i\hbar\frac{\partial|\Psi(t)\rangle}{\partial t}&=&i\hbar\left[\frac{\partial|\breve{R}_i(t)\rangle}{\partial t}\breve{R}_i^{\dagger}(t)\right]\breve{R}_i(t)|\Psi(0)\rangle\nonumber \\
&=&H(t)|\Psi(t)\rangle,
\end{eqnarray}
where $H(t)=i\hbar\frac{\partial|\breve{R}_i(t)\rangle}{\partial t}\breve{R}_i^{\dagger}(t)$
is the Hamiltonian governing the evolution of the state $|\Psi(t)\rangle$.
Thus, the Hamiltonian $H(t)$ for the Yang-Baxter system is derived through the Yang-Baxterization approach.

There are two parameters $\theta$ and $\varphi$  in the unitary
matrix $\breve{R}(\theta,\varphi)$. Consider that $\theta$ is
time-independent and $\varphi$ is time-dependent, a Hamiltonian can
be constructed from the matrix $\breve{R}(\theta,\varphi)$ as done
in Refs.\cite{J.L.K,cxg,ckg2}. The Hamiltonian is
\begin{eqnarray}\label{h0}
\hat{H}(\theta,\varphi)
=\frac{1}{\sqrt{3}}\hbar\dot{\varphi}\sin\theta \cos\theta[
2S_{i+1}^{3}(e^{-i\varphi}S_{i}^{+}S_{i+2}^{+}+e^{i\varphi}S_{i}^{-}S_{i+2}^{-})\nonumber\\
+e^{-i\varphi}(S_{i}^{+}S_{i+1}^{+}+S_{i+1}^{+}S_{i+2}^{+})
+e^{i\varphi}(S_{i}^{-}S_{i+1}^{-}+S_{i+1}^{-}S_{i+2}^{-})] \nonumber\\
+\frac{1}{3}\hbar\dot{\varphi}\cos^{2}\theta
[2(S_{i}^{3}+S_{i+1}^{3}+S_{i+2}^{3})+2S_{i+1}^{3}(S_{i}^{+}S_{i+2}^{-}+S_{i}^{-}S_{i+2}^{+})
\nonumber\\-(S_{i}^{+}S_{i+1}^{-}+S_{i+1}^{+}S_{i+2}^{-}
+S_{i}^{-}S_{i+1}^{+}+S_{i+1}^{-}S_{i+2}^{+})]
\end{eqnarray}
One can find the eigenvalues and corresponding eigenstates as follows,
\begin{eqnarray}\label{x1}
&&E_{1}=E_{2}=E_{3}=E_{4}=0 \nonumber \\
&&E_{5}=E_{6}=-\hbar\dot{\varphi}\cos\theta, \;\;E_{7}=E_{8}=\hbar\dot{\varphi}\cos\theta \nonumber \\
&&|\chi _{1}\rangle = \frac{1}{\sqrt{2}}(-|011\rangle +|110\rangle )\nonumber\\
&&|\chi _{2}\rangle = \frac{1}{\sqrt{2}}(-|001\rangle +|100\rangle )\nonumber\\
&&|\chi _{3}\rangle = \frac{1}{\sqrt{2}}(-|011\rangle +|101\rangle )\nonumber\\
&&|\chi _{4}\rangle = \frac{1}{\sqrt{2}}(|001\rangle +|010\rangle )\nonumber\\
&&|\chi _{5}\rangle = -\frac{1}{\sqrt{3}}e^{-i\varphi}\sin\frac{\theta}{2}|001\rangle
+\frac{1}{\sqrt{3}}e^{-i\varphi}\sin\frac{\theta}{2}|010\rangle
-\frac{1}{\sqrt{3}}e^{-i\varphi}\sin\frac{\theta}{2}|100\rangle +\cos\frac{\theta}{2}|111\rangle \nonumber\\
&&|\chi _{6}\rangle =
\frac{1}{\sqrt{3}}\cos\frac{\theta}{2}|001\rangle
-\frac{1}{\sqrt{3}}\cos\frac{\theta}{2}|010\rangle
+\frac{1}{\sqrt{3}}\cos\frac{\theta}{2}|100\rangle +e^{i\varphi}\sin\frac{\theta}{2}|111\rangle \nonumber\\
&&|\chi _{7}\rangle =
-e^{-i\varphi}\sin\frac{\theta}{2}|000\rangle
+\frac{1}{\sqrt{3}}\cos\frac{\theta}{2}|011\rangle
+\frac{1}{\sqrt{3}}\cos\frac{\theta}{2}|101\rangle +\frac{1}{\sqrt{3}}\cos\frac{\theta}{2}|110\rangle \nonumber\\
&&|\chi _{8}\rangle =\cos\frac{\theta}{2}|000\rangle
+\frac{1}{\sqrt{3}}e^{i\varphi}\sin\frac{\theta}{2}|011\rangle
+\frac{1}{\sqrt{3}}e^{i\varphi}\sin\frac{\theta}{2}|101\rangle +\frac{1}{\sqrt{3}}e^{i\varphi}\sin\frac{\theta}{2}|110\rangle \nonumber\\
\end{eqnarray}
It is clear that four of the eigenstates $|\chi
_{i}\rangle\;(i=1,2,3,4)$ can be decomposed to two-qubit entangled
states and single-qubit states. We get $\tau^{1,2,3,4}_{ABC}$=0
since there exists no 3-qubit entanglement for the four states. One
can easily understand that when any pair of qubit in a 3-qubit
system has maximal entanglement, 3-qubit entanglement will vanish.
According to Eq. (\ref{con}), by calculating the reduced density for
the first four eigenstates, we have $C^{1}_{AC}=1$ for the state
$|\chi _{1}\rangle$, $C^{2}_{AC}=1$ for the state $|\chi
_{2}\rangle$, $C^{3}_{AB}=1$ for the state $|\chi _{3}\rangle$ and
$C^{4}_{BC}=1$ for the state $|\chi _{4}\rangle$. One can easily
understand that when any pair of qubit in a 3-qubit system has
maximal entanglement, 3-qubit entanglement will vanish. The other
four eigenstates $|\chi _{i}\rangle\; (i=5,6,7,8)$ are 3-qubit
entangled states. These eigenstates have the same 3-tangle
$\tau_{ABC}=
\frac{16\sqrt{3}|\sin\frac{\theta}{2}\cos^{3}\frac{\theta}{2}|}{9}$.

According to Berry's theory \cite{berry}, when the parameter
$\varphi$ evolves adiabatically from 0 to 2$\pi$, the Berry phase
accumulated is,
\begin{equation}\label{r}
\gamma  = i\int^{2\pi} _{0} \langle \chi|\frac{d}{d \varphi}|\chi
\rangle d \varphi
\end{equation}
From Eq. (\ref{x1}), one can see that the Berry phase of the states $|\chi _{1}\rangle$, $|\chi
_{2}\rangle$, $|\chi _{3}\rangle$, $|\chi _{4}\rangle$ are zero. While for the other four states,
we obtain the Berry phases in the following,
\begin{eqnarray}\label{r2}
 \gamma_{5}=\gamma_{7}= \pi (1- \cos \theta)= \frac{\Omega}{2}\nonumber\\
   \gamma_{6}=\gamma_{8}=-\pi (1- \cos \theta)= -\frac{\Omega}{2}.
\end{eqnarray}
where $\Omega = 2\pi (1- \cos \theta)$ is the solid angle
enclosed by the loop on the Bloch sphere.

Introduce three operators
\begin{eqnarray}\label{i}
I_{+}=S^{+}_{1}S^{+}_{2}+S^{+}_{2}S^{+}_{3}+2S_{2}^{3}S^{+}_{1}S^{+}_{3}\nonumber\\
I_{-}=S^{-}_{1}S^{-}_{2}+S^{-}_{2}S^{-}_{3}+2S_{2}^{3}S^{-}_{1}S^{-}_{3}\nonumber\\
I_{3}=S^{3}_{1}+S^{3}_{2}+S^{3}_{3}+
    S_{2}^{3}(S^{+}_{1}S^{-}_{3}+S^{-}_{1}S^{+}_{3})\nonumber\\
    -\frac{1}{2}(S^{+}_{1}S^{-}_{2}+S^{-}_{1}S^{+}_{2}+S^{+}_{2}S^{-}_{3}+S^{-}_{2}S^{+}_{3})
\end{eqnarray}
it is not difficult to find that $(I_{\pm})^2=0$ and
$I_{3}^2=\frac{1}{4}$. We thus have a $SU(2)$ group formed by the three
operators, fulfilling conditions $[I_{+},I_{-}]=2I_{3}$ and
$[I_{3},I_{\pm}]=\pm I_{\pm}$. The Hamiltonian (\ref{h0}) can be
rewritten by using the operators
\begin{equation}\label{H}
H(\theta,\varphi)=B_{+}I_{+}+B_{-}I_{-}+B_{3}I_{3}= \mathcal
{B}\cdot \mathcal
    {J}
\end{equation}
where the parameters
$B_{+}=\frac{1}{\sqrt{3}}\hbar\dot{\varphi}\sin\theta \cos\theta
e^{-i\varphi}$,
$B_{-}=\frac{1}{\sqrt{3}}\hbar\dot{\varphi}\sin\theta \cos\theta
e^{i\varphi}$, and
$B_{3}=\frac{2}{3}\hbar\dot{\varphi}\cos^{2}\theta$. This is the
reason that Berry phase of the 3-qubit Yang-Baxter system is the
 consistence with that of the 2-qubit Yang-Baxter system \cite{J.L.K}. We also get the
same eigenvalues of Hamiltonian (\ref{h0}) as those of the 2-qubit
system \cite{J.L.K}.

\section{Summary}\label{sec4}
In summary, we have discussed a new $8\times8$ extraspecial 2-groups
operator $\mathbb{M}$ extending from the $4\times4$ extraspecial
2-groups operator $M$. A unitary $\breve{R}(\theta,\varphi)$ matrix
is constructed by the Yang-Baxterization approach. 3-qubit entangled
states can be achieved by acting the unitary
$\breve{R}(\theta,\varphi)$ matrix on the product states of 3-qubit.
Remarkably the eight states have the same degree of entanglement,
and arbitrary tripartite entangled states can be achieved depending
on the parameters $\theta$. We have constructed a Hamiltonian of 3
qubits from the unitary matrix $\breve{R}(\theta , \varphi)$ and
investigated the Berry phase of the system. The Berry phase is found
to be consistence with  that of the Yang-Baxter system of 2 qubits.
The reason is that the Hamiltonian can be rewritten by using $SU(2)$
operators.

\hspace{1cm}

The authors would like to thank Gangcheng Wang and Chunfang Sun for
useful suggestions. This work is supported by NSF of China (Grant
No. 10875026) and NUS research Grant No. WBS:
R-144-000-189-305.

\hspace{1cm}

\end{document}